# Why all the fuss about 2D semiconductors?

Andres Castellanos-Gomez

Instituto Madrileño de Estudios Avanzados en Nanociencia (IMDEA Nanoscience). 28049 Madrid, Spain.

andres.castellanos@imdea.org

The isolation of mechanically exfoliated graphene in 2004[1] sparked the research on two dimensional materials which keeps growing at a tremendous rate. It is often argued that the success of this research field is sustained by the potential of these materials to have a disruptive impact in different technological areas such as electronics and optoelectronics. However, at this stage most of the works on these systems are still focused on addressing fundamental questions and thus we might fall in the risk of selling empty promises and creating expectations in the society that might not be fulfilled in a near future. One of the real key factors behind the rise of this research field, on the other hand, is that mechanical exfoliation has democratized material science as high quality samples, showing an interesting plethora of physical phenomena, can be prepared in almost any laboratory even without specialized and expensive equipment. Just with a piece of a bulk layered material, a roll of tape and an optical microscope any trained researcher can isolate atomically thin layers of many different 2D materials ranging from wide band gap insulators to superconductors. Since 2010/2011 this relatively young field has experienced a new boost, originating from the works on the semiconducting 'cousins' of graphene, *i.e.* atomically thin, or, two dimensional, semiconductors).[2–4] Are we just entering new 'hype' phase, or are there intrinsic and profound reasons to justify this excitement of the scientific community.

One can analyze the surge of interest on 2D semiconductors as motivated by the limited success in opening a sizeable band gap in graphene. Indeed, very rapidly after the isolation of graphene, a big part of the scientific community realized that although graphene shows a remarkably high carrier mobility, is very flexible (and tough) and almost transparent, its lack of bandgap can be a severe handicap for its use in certain applications. Therefore, from 2007/2008 great efforts were made to open a sizeable gap in graphene by patterning graphene into nanoribbons,[5] hydrogenation [6] or by applying a perpendicular electric field,[7] but without achieving a breakthrough that guaranteed its applicability. Another part of the community, on the other hand, focused on finding semiconducting counterparts to graphene rather than modifying graphene to open a gap.

To date, more than ten different 2D semiconductors (with band gap values spanning from few meV up to several eV) have been experimentally isolated and there are potentially hundreds that could be isolated in a near future. Because of this broad catalogue of materials, it is always possible to find a 2D semiconductor optimal for a certain application (see Figure 1). Moreover, most of the 2D semiconductor families studied to date have also shown interesting phenomena, some of them observed for the first time on these 2D systems. In the following the most relevant families of 2D semiconductors will be introduced, stressing on the aspects that make them especially interesting.

**<u>Transition metal dichalcogenides:</u>**

With a general formula $MX_2$ (M being a transition metal and X a chalcogen) this family of 2D materials is, after graphene, probably the most studied one. These materials present an intrinsic bandgap within the visible part of the spectrum. Due to quantum confinement in the out of plane direction, the bandgap strongly changes with the number of layers and a transition from direct gap to indirect gap has been observed when the number of layers is increased from monolayer to multilayers.[2,3] Another interesting feature of these materials is that the large spin-orbit interaction in these compounds, due to the heavy transition metals, leads to a splitting of the valence band that strongly affects their optical spectra.

It is also worth mentioning that these materials present a huge exciton binding energy, which makes them ideal to study excitonic physics through photoluminescence measurements, even at room temperature. These studies unraveled interesting phenomena such as the generation of charged excitons (trions)[8] or the valley polarized photoluminescence emission when one valley is optically pumped with circularly polarized light.[9–11] The recent observations of single-photon emitters in $WSe_2$ with very narrow emission linewidth (~100 µeV), due to localized excitonic states that are related to defects,[12–15] have attracted the attention of the optical spectroscopy community on this material.

**Hexagonal boron nitride:**

This material has been traditionally considered as a good substrate or encapsulation layer to fabricate nanodevices with other 2D materials.[16] However, since 2014 hexagonal boron nitride has also gained quite some attention in the photonics community because of the recent demonstration of strongly confined phonon-polariton modes (collective oscillations, with frequencies typically in the mid-IR wavelength range, resulting from the coupling of light photons with optical phonons in polar dielectrics). Using scanning near-field optical microscopy polaritonic waves were launched, detected, and imaged in real space in thin boron nitride flakes and their measured dispersion exhibited hyperbolic dependencedispersion.[17,18] This strong confinement of radiation arises from the anisotropy in the permittivity tensor in boron nitride, whose in-plane and out-of-plane components have a different sign. These peculiar materials are known as hyperbolic materials and so far they have been mainly fabricated artificially with nanofabrication techniques. Boron nitride is a natural hyperbolic material which excels the man-made hyperbolic materials which tend to suffer from high losses yielding short propagation lengths and broadband resonances. Moreover, while man-made hyperbolic materials have shown confinement values of only $\lambda/12$ with poor quality factors of ~5, boron nitride nanocones have recently shown strongly three-dimensionally confined 'hyperbolic polaritons' (confinements of up to $\lambda/86$) and exhibit high-quality factors ($Q$ up to 283).[19] Therefore, these phonon-polariton modes in boron nitride can have a strong impact in nanophotonics to confine radiation to a very small length scale (sub-diffraction limit) which eventually can result in new imaging applications in the mid-IR part of the spectrum.[20]

**Black phosphorus and other quasi 1D two dimensional materials:**

Despite the youth of this 2D material (the first works on atomically thin black phosphorus were reported just two years ago [21,22]) the number of studies on black phosphorus is growing rapidly. The bloom of interest on this material can be due to the combination of several factors: it shows one of the highest charge carrier mobility reported for 2D semiconductors (typically 100 – 1000 cm$^2$/V·s), its band gap spans over a wide range of the electromagnetic spectrum (from mid-IR to visible) and it shows rather exotic in-plane anisotropy (most 2D semiconductors have a marked anisotropy between the in-plane and out-of-plane directions but they are typically rather isotropic within the basal plane).

Regarding the black phosphorus in-plane anisotropy, unlike in graphite (where carbon atoms bond with three neighboring atoms through sp$^2$ hybridized orbitals) in black phosphorus each phosphorus atom bonds to three neighboring phosphorus atoms through sp$^3$ hybridized orbitals, causing the phosphorus atoms to be arranged in a puckered honeycomb lattice formation.[23] This structure is the seed of an anisotropic band structure which leads to highly anisotropic electrical, thermal, mechanical and optical properties. This is in striking contrast to graphene, boron nitride or Mo- and W- based transition metal dichalcogenides that they do not present noticeable in-plane anisotropy. Regarding its anisotropic optical properties, black phosphorus has shown a marked linear dichroism,[23] optical absorption that depends on the relative orientation between the materials lattice and an incident linearly polarized light. The dichroism has strong implications for its Raman spectra, plasmonic and screening effects and photoresponse. [24,25] Recently, it has been also demonstrated how the photoluminescence yield of black phosphorus always shows a high degree of polarization along the armchair direction of the flake,[25] opening the door for on-chip light polarization manipulation.

Similar to black phosphorus, other 2D semiconductors (Re- based chalcogenides [26] and trichalcogenides such as $TiS_3$ [27]) also demonstrated marked electrical and optical in-plane anisotropy.

**Silicene and other group IV cousins of graphene:**

Just below carbon, in the group IV of the periodic table we find other elements that (similar to graphene) can also form elemental 2D materials based on silicon (silicene), germanium (germanene) and tin (stanene) atoms, although adopting a buckled structure.[28] Due to the larger spin-orbit coupling of silicon, germanium and tin with respect to carbon the calculated band structure of these 2D elemental materials differ from that of graphene and a band gap should open (~2 meV for silicene, ~20 meV for germanene and ~300 meV for stanene).[29]

So far these narrow gap 2D semiconductors have been epitaxially grown on metallic surfaces [30–32] in ultrahigh vacuum and their limited environmental stability hampers to study their properties with ex-situ techniques. Moreover, it has been shown that the metallic substrates employed for the epitaxial growth strongly modify the electronic properties of the silicene/germanene/stanene. Therefore, experimentally studying the intrinsic properties of these 2D semiconductors (and exploiting them in applications such as mid-IR and NIR optoelectronics) seemed pretty hopeless. Nonetheless, Tao *et al*.[33] have implemented a method to encapsulate silicene, to take it out of the ultrahigh vacuum environment and to fabricate electronic devices out of it. This route to overcome the initial pessimism about the environmental instability of this family of 2D materials opens the door to further study them, especially in the context of optoelectronics and photonics where there is still a total lack of experimental works.

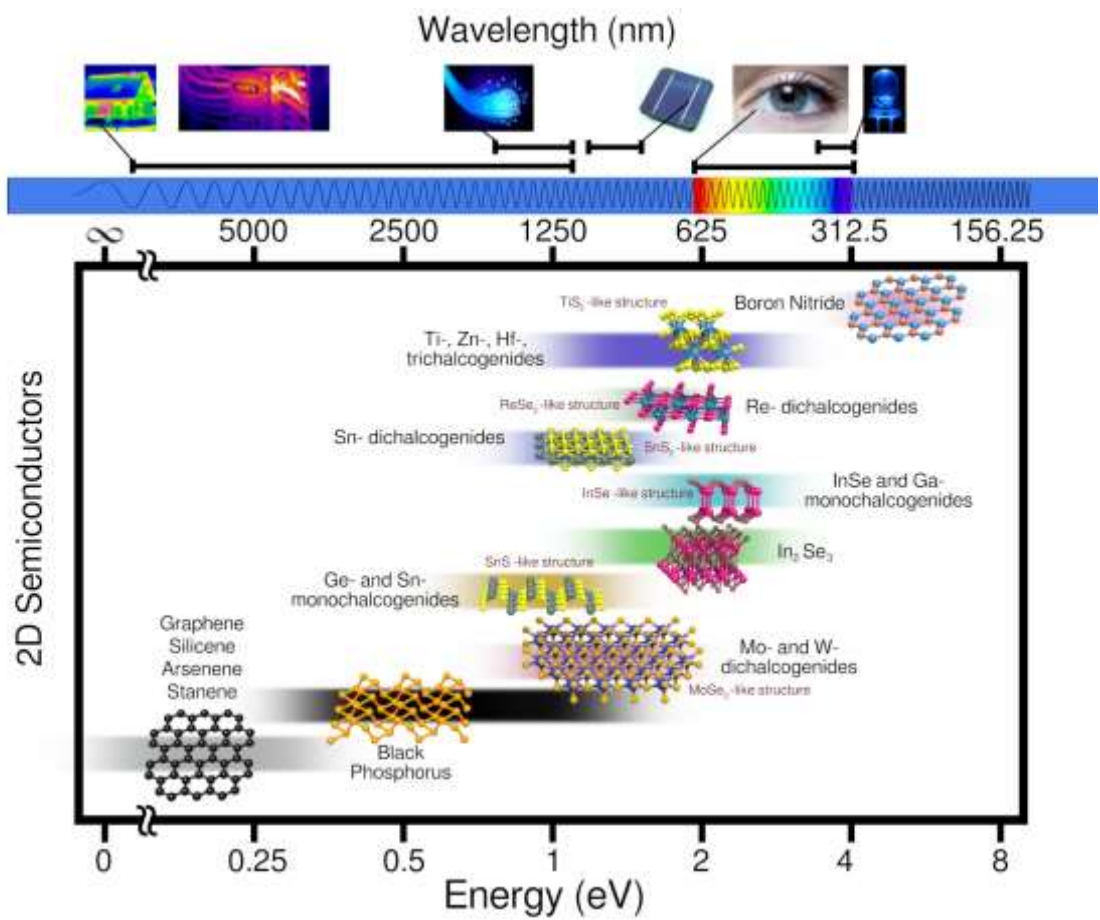

**Figure 1:** Comparison of the band gap values for different 2D semiconductor materials families studied so far. The crystal structure is also displayed to highlight the similarities and differences between the different families. The horizontal degraded bars indicate the range of band-gap values that can be spanned by changing the number of layers, straining or alloying.

| Material | Mobility | Gap | Exciton binding energy | Plasma frequency | Observed phenomena |
|---|---|---|---|---|---|
| $MoS_2$ | 1-100 cm$^2$/V·s [4] | 1.3-1.9 eV [2,3] | ~400 meV [34,35] | ~25 meV [36] | Charged excitons (trions) [8]<br>Valley polarization [9–11]<br>Strong spin-orbit interaction<br>Quantum confinement [2,3] |
| BP | 100-1000 cm$^2$/V·s [21] | 0.3-1.5 eV | ~400 meV [37] | ~400 meV [38] | Anisotropic photoluminescence [25]<br>Quantum confinement |
| hBN | | 5.9 eV [39] | ~150 meV [39] | ~8 eV [40] | Hyperbolic material [17,18]<br>Confined phonon-polariton modes [19,20] |
| Silicene | 100 cm$^2$/V·s [33] | 2 meV [29] | | | Fairly unexplored [28] |

**Table 1:** Summary of properties of different properties of 2D semiconductors relevant for optoelectronics and photonics applications.

**Acknowledgements**

AC-G acknowledges Enrique Sahagun from SCIXEL for his support in the elaboration of the graphical material. AC-G also acknowledges financial support from the BBVA Foundation through the fellowship "I Convocatoria de Ayudas Fundacion BBVA a Investigadores, Innovadores y Creadores Culturales" ("Semiconductores ultradelgados: hacia la optoelectronica flexible"), from the MINECO (Ramón y Cajal 2014 program, RYC-2014-01406), from the MICINN (MAT2014-58399-JIN) and from the Comunidad de Madrid (MAD2D-CM Program (S2013/MIT-3007)).